\documentclass[twocolumn,showpacs,aps,prd]{revtex4-1}
\usepackage{epsfig}
\usepackage{graphicx}
\usepackage{dcolumn}
\usepackage{bm}
\usepackage{ltablex,booktabs}
\usepackage{overpic}
\usepackage{subfigure}
\usepackage{float}
\usepackage{color}
\usepackage{amsmath}
\usepackage{mathcomp}
\usepackage{mathrsfs}
\usepackage{multirow}
\usepackage{rotating}
\usepackage{amssymb}
\usepackage{gensymb}
\usepackage{amsmath}
\usepackage{tabularx}
\usepackage{tikz-feynman}
\usepackage{lineno}
\usepackage[bookmarksnumbered, pdfstartview=FitH,colorlinks,urlcolor=blue, citecolor=blue,linkcolor=blue] {hyperref}
\usepackage{epstopdf}

\begin{document}
\normalsize
\parskip=5pt plus 1pt minus 1pt
\title{\boldmath Search for the leptonic decays $D^{*+}\to e^+\nu_e$ and $D^{*+}\to \mu^+\nu_\mu$}

\author{
M.~Ablikim$^{1}$, M.~N.~Achasov$^{12,b}$, P.~Adlarson$^{72}$, M.~Albrecht$^{4}$, R.~Aliberti$^{33}$, A.~Amoroso$^{71A,71C}$, M.~R.~An$^{37}$, Q.~An$^{68,55}$, Y.~Bai$^{54}$, O.~Bakina$^{34}$
% ...
\\
\vspace{0.2cm}
(BESIII Collaboration)\\
\vspace{0.2cm} {\it
$^{1}$ Institute of High Energy Physics, Beijing 100049, People's Republic of China\\
% ...
$^{4}$ Bochum  Ruhr-University, D-44780 Bochum, Germany\\
% ...
$^{12}$ G.I. Budker Institute of Nuclear Physics SB RAS (BINP), Novosibirsk 630090, Russia\\
% ...
$^{33}$ Johannes Gutenberg University of Mainz, Johann-Joachim-Becher-Weg 45, D-55099 Mainz, Germany\\
$^{34}$ Joint Institute for Nuclear Research, 141980 Dubna, Moscow region, Russia\\
% ...
$^{37}$ Liaoning Normal University, Dalian 116029, People's Republic of China\\
% ...
$^{54}$ Southeast University, Nanjing 211100, People's Republic of China\\
$^{55}$ State Key Laboratory of Particle Detection and Electronics, Beijing 100049, Hefei 230026, People's Republic of China\\
% ...
$^{68}$ University of Science and Technology of China, Hefei 230026, People's Republic of China\\
% ...
$^{71}$ University of Turin and INFN, (A)University of Turin, I-10125, Turin, Italy; (B)University of Eastern Piedmont, I-15121, Alessandria, Italy; (C)INFN, I-10125, Turin, Italy\\
$^{72}$ Uppsala University, Box 516, SE-75120 Uppsala, Sweden\\
% ...
$^{b}$ Also at the Novosibirsk State University, Novosibirsk, 630090, Russia\\
% ...
}
}

\begin{abstract}
  We present the first search for the leptonic decays $D^{*+}\to e^+\nu_e$ and
  $D^{*+}\to \mu^+\nu_\mu$ by analyzing a data sample of electron-positron
  collisions recorded with the BESIII detector at center-of-mass energies
  between 4.178 and 4.226~GeV, corresponding to an integrated luminosity of
  6.32~fb$^{-1}$. No significant signal is observed. The upper limits on the branching fractions for
  $D^{*+}\to e^+\nu_e$ and $D^{*+}\to \mu^+\nu_\mu$
  are set to be $1.1 \times 10^{-5}$ and $4.3 \times 10^{-6}$ at 90\% confidence level, respectively.
\end{abstract}
\maketitle

\section{Introduction}
Experimentally, the study of pure leptonic decays of ground state charm mesons has entered the stage of precision measurement~\cite{lep}. However, further exploration is needed for the pure leptonic decays of excited pseudoscalar mesons. The $D^{*+}$ meson is the lightest excited state of the $D^{+}$ meson, whose quark
content is $c\bar{d}$, with a spin equal to 1. 

\begin{figure}[h]
\centering
\includegraphics[height=2.8cm,width=8cm]{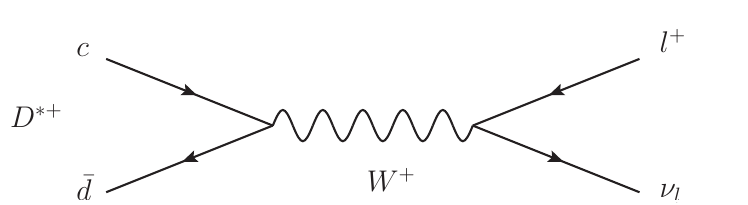}
    \caption{Feynman diagram of $D^{*+}\to \ell^+\nu_\ell$.}
  \label{fig:lep}
\end{figure}

In the Standard Model (SM), the
leptonic decays $D^{*+}\to \ell^+\nu_\ell$~($\ell= e, \mu$) are described by
the annihilation of the initial quark-antiquark pair into a virtual $W^+$ that
materializes $\ell\nu_\ell$ pair as shown in Figure~\ref{fig:lep}. Throughout
this paper, charge conjugation is always implied, unless explicitly specified otherwise.

The examination of $D^{*+}\to \ell^+\nu_\ell$ decays holds significance, as their decay rates are linked to the Cobibbo-Kobayashi-Maskawa (CKM) matrix elements. 
The decay width of $D^{*+}\to \ell^+\nu_\ell$ can be parameterized by the
$D^{*+}$ decay constant $f_{D^{*}}$~\cite{decayrate, Akeroyd1:2021YL} via
\begin{equation}
  \Gamma= \frac{G_F^2}{12\pi}|V_{cd}|^2f^2_{D^{*}}
  m_{D^{*}}^3 \left(1-\frac{m_\ell^2}{m_{D^{*}}^2}\right)^2 \left(1+\frac{m_\ell^2}{2m_{D^{*}}^2}\right),
  \label{deacayrate}
\end{equation}
where $G_F$ is the Fermi coupling constant, $|V_{cd}|$ is the CKM matrix element between the $c$ and $d$ quarks, and $m_\ell$
($m_{D^{*}}$) is the mass of the lepton ($D^{*+}$). 
Several theoretical studies have anticipated the $f_{D^{*}}$, such as nonrelativistic quark model, relativistic quark model, lattice QCD, etc., predicting $f_{D^{*}}$ values between 186~MeV and 391~MeV~\cite{arxiv}. In addition, as the $D^{*+}$ can decay by the strong interaction, the branching fractions of these weak
decays are at the $10^{-10}$ level within the SM~\cite{Akeroyd1:2021YL, Akeroyd1:2017ND, Akeroyd1:2001KB}, but potential contributions from new
pseudoscalar interactions beyond the SM could lift the branching fractions
up to $10^{-5}$~\cite{Akeroyd:2009tn, Akeroyd:2002pi, Akeroyd:2003jb}. Any observation of the leptonic $D^{*+}$ decays at a rate above the SM prediction would be a potential hint of new physics.

This paper reports the first searches ever performed for the leptonic decays $D^{*+} \to e^+\nu_e$ and
$D^{*+}\to \mu^+\nu_\mu$, using a data sample corresponding to 
an integrated luminosity of 6.32~fb$^{-1}$,
recorded by the BESIII detector at center-of-mass (CM) energies~($\sqrt{s}$) ranging from 4.178 to 4.226~GeV.

\section{DETECTOR and DATA SETS} \label{sec:detector_dataset}
The BESIII detector~\cite{Ablikim:2009aa} records symmetric $e^+e^-$ collisions
provided by the BEPCII storage ring~\cite{Yu:IPAC2016-TUYA01}, which operates in $\sqrt{s}$ from 2.00 to 4.95~GeV, with a peak luminosity of $1\times10^{33}$~cm$^{-2}$s$^{-1}$ achieved at $\sqrt{s}$ = 3.77~GeV. BESIII has collected large
data samples in this energy region~\cite{Ablikim:2019hff}. The cylindrical core
of the BESIII detector covers 93\% of the full solid angle and consists of a
helium-based multilayer drift chamber~(MDC), a plastic scintillator
time-of-flight system~(TOF), and a CsI(Tl) electromagnetic calorimeter~(EMC),
which are all enclosed in a superconducting solenoidal magnet providing a 1.0~T
magnetic field. The solenoid is supported by an octagonal flux-return yoke with
resistive plate counter muon identification modules interleaved with steel~(MUC)~\cite{muon}.
The charged-particle momentum resolution at $1~{\rm GeV}/c$ is $0.5\%$, and the specific ionization energy loss (d$E$/d$x$) resolution is $6\%$ for electrons from Bhabha scattering.
The EMC measures photon energies with a resolution of $2.5\%$ ($5\%$) at
$1$~GeV in the barrel (end-cap) region. The time resolution in the TOF barrel
region is 68~ps, while that in the end-cap region is 110~ps. The end-cap TOF
system was upgraded in 2015 using multigap resistive plate chamber technology,
providing a time resolution of 60~ps~\cite{etof}.

The data samples used in this analysis are listed in Table~\ref{energe} and provide
a large sample of $D^{*\pm}$ mesons from $e^+e^-\to D^{*+}D^{*-}$ events
with a cross section of about 3~nb~\cite{Abe:2006fj}. 

\begin{table}[htb]
 \renewcommand\arraystretch{1.25}
 \centering
 \caption{Integrated luminosities $\mathcal{L}_{\rm int}$~\cite{ref:lumino1, ref:lum} of the data samples at different energies. The first and
   second uncertainties are statistical and systematic, respectively.}
 \begin{tabular}{cc}
 \hline 
 $\sqrt{s}$ (GeV) & $\mathcal{L}_{\rm int}$ (pb$^{-1}$)\\
 \hline
  4.178 &$3189.0\pm0.2\pm31.9$\\
  4.189 &$526.7\pm0.1\pm2.2$ \\
  4.199 &$526.0\pm0.1\pm2.1$ \\
  4.209 &$517.1\pm0.1\pm1.8$ \\
  4.219 &$514.6\pm0.1\pm1.8$ \\
  4.226 &$1056.4\pm0.1\pm7.0$ \\
  \hline
 \end{tabular}
 \label{energe}
\end{table}

Simulated Monte Carlo~(MC) samples produced with {\sc geant4}-based~\cite{GEANT4}
software, which includes the geometric description of the BESIII detector and
the detector response, are used to determine the detection efficiency and to
estimate the background contributions. The simulation includes the beam-energy
spread and initial-state radiation~(ISR) in the $e^+e^-$ annihilation modeled
with the generator {\sc kkmc}~\cite{KKMC}. Inclusive MC samples of 40 times the size of the data samples are used to simulate the background contributions. The
inclusive MC samples contain no signal decays and include the production of
open-charm processes, the ISR production of vector charmonium(-like) states,
and the continuum processes incorporated in {\sc kkmc}~\cite{KKMC}. The known decay modes
are modeled with {\sc evtgen}~\cite{EVTGEN} using world averaged branching fraction
values~\cite{PDG}, and the remaining unknown decays from the charmonium states
with {\sc lundcharm}~\cite{LUNDCHARM}. Final-state radiation from charged
final-state particles is incorporated with {\sc photos}~\cite{PHOTOS}.

\section{DATA ANALYSIS}
\label{chap:event_selection}
 The single tag~(ST) sample comprises events in which only the $D^{*-}$ meson is reconstructed via hadronic
decay modes.  The double tag~(DT) sample consists of events in which
the leptonic decays of $D^{*+}\to \ell^+\nu_\ell$ are reconstructed in the
systems recoiling against the ST candidates~\cite{MarkIII-tag}. The branching
fraction for $D^{*+}\to \ell^+\nu_\ell$ can be determined by
\begin{equation}
{\mathcal B}(D^{*+}\to \ell^+\nu_\ell) = \frac{N^{\rm tot}_{\rm DT}}{N^{\rm tot}_{\rm ST}\bar \epsilon_{D^{*+}\to \ell^+\nu_\ell}},
\label{eq:brofdt}
\end{equation}
where $N^{\rm tot}_{\rm ST}$ and $N^{\rm tot}_{\rm DT}$ are the ST and DT yields in the data sample, respectively, and $\bar \epsilon_{D^{*+}\to \ell^+\nu_\ell} = \sum_i (N^i_{\rm ST} \epsilon^i_{\rm DT}/f\epsilon^i_{\rm ST})/N^{\rm tot}_{\rm ST}$
is the effective averaged efficiency of reconstructing the $D^{*+}\to \ell^+\nu_\ell$
decay weighted by the ST yields, where $i$ indicates the $i^{\rm th}$ tag mode. Here, $\epsilon^i_{\rm ST}$ and $\epsilon^i_{\rm DT}$ are the efficiencies of reconstructing the ST and DT candidates in the $i^{\rm th}$ tag mode (called the ST efficiency and DT efficiency respectively). The ST efficiency is the average efficiency of ST $D^{*-}$ from  $D^{*+}D^{*-}$, $D^{*-}D^{+}$, $D^{0}D^{*-}\pi^{+}$, $D^{+}D^{*-}\pi^{0}$. $f$ represents the ratio of the number of $D^{*-}$ contributed by all processes to that by $D^{*+}D^{*-}$ pairs, $f = (2\sigma_{D^{*-}D^{*+}}+\sigma_{D^{*-}D^{+}}+\sigma_{D^{0}D^{*-}\pi^{+}}+\sigma_{D^{+}D^{*-}\pi^{0}})/2\sigma_{D^{*+}D^{*-}}$, where the $\sigma$ denotes the corresponding cross section~\cite{Akeroyd:dd, Akeroyd:ddpi}. Note that, the cross section of $D^{+}D^{*-}\pi^{0}$ is set to be half of $D^{0}D^{*-}\pi^{+}$, according to the isospin symmetry.

\subsection{Single-tag analysis}
The ST candidates are reconstructed through the two main decay modes of $D^{*-}$
mesons: $\bar{D}^0 \pi^-$ and $D^- \pi^0$. To reconstruct $\bar D^0$ mesons,
three hadronic decay modes, $K^+\pi^-$, $K^+\pi^-\pi^0$, and
$K^+\pi^-\pi^+\pi^-$, are used. To reconstruct $D^-$ mesons, six hadronic
decay modes, $K^+\pi^-\pi^-$, $K^+K^-\pi^-$, $K^+\pi^-\pi^-\pi^0$,
$K_S^0\pi^-$, $K_S^0\pi^-\pi^0$, and $K^0_S \pi^+\pi^-\pi^-$ are used.


All charged tracks are required to
be within the polar angle ($\theta$) range of $|\rm{cos\theta}|<0.93$, where
$\theta$ is defined with respect to the $z$ axis, which is the symmetry axis of
the MDC. For charged tracks not originating from $K_S^0$ decays, the distance
of closest approach to the interaction point (IP) must be less than 10\,cm
along the $z$ axis, $|V_{z}|$, and less than 1\,cm in the transverse plane,
$|V_{xy}|$. Particle identification~(PID) for charged tracks combines
measurements of the d$E$/d$x$ and TOF to form likelihoods $\mathcal{L}(h)~(h=K,\pi)$ for each hadron $h$
hypothesis. Tracks are identified as charged kaons and pions by comparing the
likelihoods for the kaon and pion hypotheses, $\mathcal{L}(K)>\mathcal{L}(\pi)$
and $\mathcal{L}(\pi)>\mathcal{L}(K)$, respectively.

Each $K_{S}^0$ candidate is reconstructed from two oppositely charged tracks
satisfying $|V_{z}|<$ 20~cm. The two charged tracks are assigned the charged pion hypothesis without imposing any PID requirements. The candidates pions are constrained to
originate from a common vertex and are required to have an invariant mass
within $|M_{\pi^{+}\pi^{-}} - m_{K_{S}^{0}}|<$ 12~MeV$/c^{2}$, where
$m_{K_{S}^{0}}$ is the known $K^0_{S}$ mass~\cite{PDG}. The decay length of
the $K^0_S$ candidate is required to be twice greater than its uncertainty.

 Photon candidates are identified using showers in the EMC. The deposited
energy of each shower must be more than 25~MeV in the barrel region
($|\!\cos\theta|< 0.80$) and more than 50~MeV in the end-cap region
($0.86 <|\!\cos\theta|< 0.92$). To exclude showers associated with charged
tracks, the angle subtended by the EMC shower and the position of the closest
charged track at the EMC must be greater than 10 degrees as measured from the
IP. To suppress electronic noise and showers unrelated to the event, the
difference between the EMC time and the event start time is required to be
within [0, 700]\,ns.

The $\pi^0$ candidates are reconstructed through  $\pi^0\to \gamma\gamma$
decays. The diphoton invariant masses $M_{\gamma\gamma}$ must lie within the range 
$[0.115, 0.150]$~GeV/$c^{2}$. For diphoton combinations satisfying this requirement $M_{\gamma\gamma}$ is kinematically constrained to the known $\pi^{0}$ mass~\cite{PDG}.



If there are multiple ST candidates found, the candidate with the minimal $|\Delta E| \equiv |E_{D^{*-}}-E_{\rm beam}|$ is selected, where $E_{\rm beam}$ is the beam energy and
$E_{D^{*-}}$ is the reconstructed energy of the ST candidate in the $e^+e^-$ CM frame. Furthermore,
it is possible that there are multiple combinations having exactly the same $\Delta E$ value
for the ST modes with $\pi^-(\pi^0)$ as final states of $\bar{D}^0(D^-)$ 
decays. Specifically, in the case of $D^{*-}\to \pi^{-} \bar{D}^0(D^{*-} \to \pi^0 D^-)$ with $\bar{D}^0\to X\pi^-(D^-\to X\pi^0)$, the value of  $\Delta E$  remains unchanged even if the two $\pi^-(\pi^0)$ is switched. 
In this case, the least 
$|\Delta M| \equiv |M_{D^{*}} - m_{D^{-(0)}}|$ is used to identify the $\pi^{-(0)}$ meson produced from $D^{*-}$ decays ($\pi^{-(0)}_{D^*}$) for further analysis.
Here,
$M_{D^{*}}$ is the reconstructed mass of $D^{*-}$ candidate and
$m_{D^{-(0)}}$ is the nominal mass of the $D^{-(0)}$ meson~\cite{PDG}.


The $\pi^{-(0)}_{D^*}$ candidate is required to have momentum less than 100~MeV/$c$
to suppress backgrounds. Since the CM energies are about 160-210~MeV higher than the $D^{*+}D^{*-}$ mass threshold, the $D^{*-}$ is boosted. MC studies show that more than 90\% of signals have $\cos\theta_{D\pi} > 0$, where $\theta_{D\pi}$ is the opening angle between $D^{0(-)}$ and $\pi^{-(0)}_{D^*}$ in the $e^+e^-$ CM frame. Therefore, a requirement of  $\cos\theta_{D\pi} > 0$ is applied to suppress backgrounds caused by the misreconstruction of the 
$\pi^{-(0)}_{D^*}$.

The variables $\Delta E$, the reconstructed mass of $\bar{D}^{0}$($D^{-}$) candidate ($M_{D}$), and the beam-constrained mass, $M_{\rm BC} = \sqrt {E_{\rm beam}^2-|\vec {p}_{D^{*-}}|^2}$, where $\vec {p}_{D^{*-}}$ is the reconstructed $D^{*-}$ momentum in the $e^+e^-$ CM frame, are used to further suppress combinatorial backgrounds. Requirements on these variables are listed in Table~\ref{tab:stcri}, which correspond to 3$\sigma$ regions for the signal process. Here $\sigma$ stands for the standard deviations for the variables that are determined by the fits on the corresponding distributions for each tag. 

\begin{table*}[!htp]
 \renewcommand\arraystretch{1.25}
  \centering
  \caption{ST selection equirements on $\Delta E$, $M_{D}$, and $M_{\rm BC}$ for each tag modes.}
\label{tab:stcri}
\begin{tabular}{lccc}
\hline
Tag mode                            & $\Delta E$~(GeV) & $M_{D}$~(GeV/$c^2$) & $M_{\rm BC}$~(GeV/$c^2$) \\ \hline
$\bar{D}^{0}\to K^+\pi^-$           & $(-0.024, 0.028)$   & (1.846, 1.885)        & (2.001, 2.024)\\
$\bar{D}^{0}\to K^+\pi^-\pi^+\pi^-$ & $(-0.024, 0.030)$   & (1.850, 1.880)        & (1.999, 2.025)\\
$\bar{D}^{0}\to K^+\pi^-\pi^0$      & $(-0.031, 0.047)$   & (1.832, 1.890)        & (1.994, 2.030)\\
$D^- \to K^+\pi^-\pi^-$             & $(-0.025, 0.032)$   & (1.854, 1.886)        & (1.997, 2.029)\\
$D^- \to K^+K^-\pi^-$               & $(-0.025, 0.032)$   & (1.856, 1.884)        & (2.000, 2.024)\\
$D^- \to K^+\pi^-\pi^-\pi^0$        & $(-0.032, 0.046)$   & (1.838, 1.893)        & (1.998, 2.026)\\
$D^- \to K_S^0\pi^-$                & $(-0.025, 0.028)$   & (1.852, 1.890)        & (1.998, 2.027)\\
$D^- \to K_S^0\pi^-\pi^0$           & $(-0.036, 0.052)$   & (1.831, 1.900)        & (1.992, 2.033)\\
$D^- \to K_S^0\pi^-\pi^+\pi^-$      & $(-0.027, 0.032)$   & (1.852, 1.887)        & (1.999, 2.025)\\ \hline
\end{tabular}
\end{table*}

\begin{figure}[!htp]
  \centering
  \includegraphics[width=0.48\textwidth]{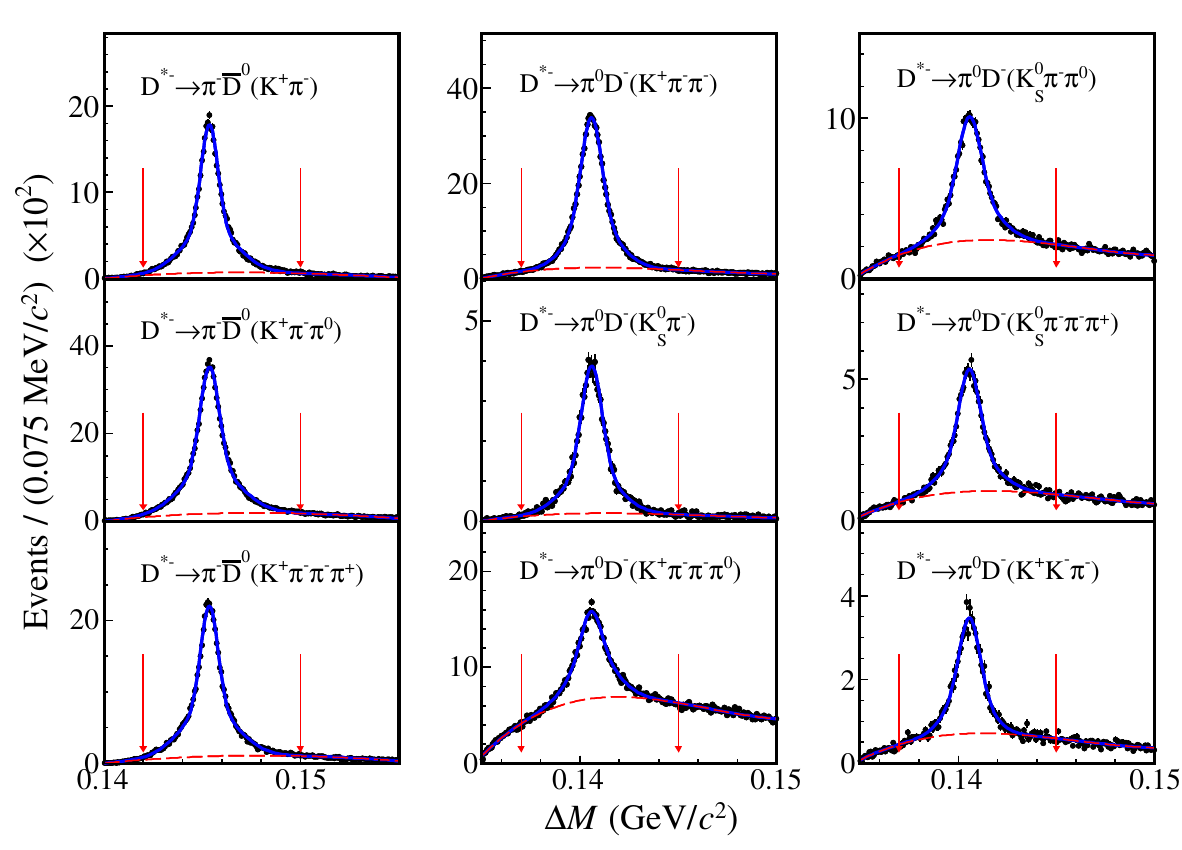}
  \caption{Fits to the $\Delta M$ distributions of the accepted ST $D^{*-}$
    candidates at $\sqrt{s}=4.178$~GeV. The points with error bars are data. The blue
    solid curves and red dashed curves represent the best fits and fitted
    combinatorial backgrounds, respectively. The pairs of red arrows indicate
    the $\Delta M$ signal region.}
  \label{fig:styeild}
\end{figure}

To determine the ST signal yield, a maximum-likelihood fit to the
distributions of $\Delta M$ is
performed for each tag mode at each energy point. In the fit, the signal shape is
described by a double-Gaussian function and the background shape by a third-order polynomial. The ST yields~($N^{i}_{\rm ST}$) and
efficiencies~$\epsilon^i_{\rm ST}$ are extracted from the fits to data and
inclusive MC samples, respectively. Signal regions of $\Delta M$ are set at three times the resolution.  Figure~\ref{fig:styeild} shows the fit results at 4.178~GeV and Table~\ref{tab:steff} summarizes the signal regions of
$\Delta M$, ST signal yields, and ST efficiencies at 4.178~GeV. The fit results at
other energy points can be found at Appendix~\ref{app:ST}. These fits give a total ST yield of $N_{\rm ST}^{\rm tot}$ = $516256\pm1870$ at 4.178-4.226~GeV.

\begin{table*}[htbp]
  \renewcommand\arraystretch{1.25}
  \centering
  \caption{ST yields and efficiencies for each tag mode at 4.178~GeV.
    Uncertainties are statistical only.}
  \label{tab:steff}
  \begin{tabular}{lccc}
    \hline
    Tag mode                            & $\Delta M$~(GeV/$c^2$)  & $N^{i}_{\rm ST}$ & $\epsilon^i_{\rm ST}$~(\%)\\ \hline
    $\bar{D}^{0} \to K^+\pi^-$          & (0.142, 0.150)           & $40325\pm453$    & $8.14\pm0.04$ \\
    $\bar{D}^{0}\to K^+\pi^-\pi^+\pi^-$ & (0.142, 0.150)           & $40325\pm453$    & $4.65\pm0.02$ \\
    $\bar{D}^{0}\to K^+\pi^-\pi^0$      & (0.142, 0.150)           & $86568\pm888$    & $4.86\pm0.02$ \\
    $D^- \to K^+\pi^-\pi^-$             & (0.137, 0.145)           & $76303\pm511$    & $14.87\pm0.05$\\
    $D^- \to K^+K^-\pi^-$               & (0.137, 0.145)           & $6389\pm241$     & $11.66\pm0.13$ \\
    $D^- \to K^+\pi^-\pi^-\pi^0$        & (0.137, 0.145)           & $24411\pm648$    & $6.68\pm0.04$ \\
    $D^- \to K_S^0\pi^-$                & (0.137, 0.145)           & $9094\pm193$     & $14.85\pm0.15$\\
    $D^- \to K_S^0\pi^-\pi^0$           & (0.137, 0.145)           & $20942\pm561$    & $7.95\pm0.05$ \\
    $D^- \to K_S^0\pi^-\pi^+\pi^-$      & (0.137, 0.145)           & $10922\pm225$    & $8.65\pm0.08$ \\ \hline
    Total                               &                         & $326405\pm1567$  &               \\ \hline
\end{tabular}
\end{table*}

\subsection{Double-tag analysis}
To search for the $D^{*+}\to\ell^+\nu_\ell$ decays recoiling against  ST
candidates, we require that there is only one charged track (the number of extra charged tracks, $N_{\rm extra}^{\rm charge}$, is zero), 
which is identified as an $e^+$ or a $\mu^+$,
and the maximum energy of photon(s) not used in the ST candidate selection, $E_{\rm{extra}~\gamma}^{\rm max}$, must
be less than 0.3~GeV, 
 to suppress
backgrounds associated with photon(s).

Measurements in the MDC, TOF and EMC are used to construct 
combined likelihoods ($\mathcal{L}'$) under the positron, pion, and kaon
hypotheses. Positron candidates are required to satisfy
$\mathcal{L}'(e)>0.001$ and
$\mathcal{L}'(e)/(\mathcal{L}'(e)+\mathcal{L}'(\pi)+\mathcal{L}'(K))>0.8$. To
reduce background from hadrons and muons, the positron candidate is further
required to have a deposited energy in the EMC greater than 80\% of its
momentum as determined from its trajectory in the MDC.  Muon PID uses information from  the
EMC and MUC. The energy deposited in the EMC is required to be in the range of
$(0.0,0.3)$~GeV. The muon penetrates further than other charged particles, and thus has a deeper hit depth in the MUC. The penetrating
depths of muon candidates are required to satisfy the criteria listed in
Table~\ref{tab:muonid}.

\begin{table}[hbtp]
 \renewcommand\arraystretch{1.25}
  \centering
\begin{center}
\caption{Requirements of the hit depth in the MUC for muon candidate.}
\begin{tabular}{ccc} \hline
$|\!\cos\theta|$ & $\ \ \ \ \ p$ (GeV/$c$)\ \ \    & Depth (cm) \\ \hline
               & $\ \ \ \ \ p\le0.88$\ \ \     & $\ \ \ >17.0\ $        \\
\ (0.00, 0.20)    & $\ \ \ \ \ 0.88<p<1.04$\ \ \  & $\ \ \ >100.0\times p-71.0\ $ \\
               & $\ \ \ \ \ p\ge1.04$\ \ \     & $\ \ \ >33.0\ $ \\ \hline
               & $\ \ \ \ \ p\le0.91$\ \ \     & $\ \ \ >17.0\ $        \\
\ (0.20, 0.40)    & $\ \ \ \ \ 0.91<p<1.07$\ \ \  & $\ \ \ >100.0\times p-74.0\ $ \\
               & $\ \ \ \ \ p\ge1.07$\ \ \     & $\ \ \ >33.0\ $ \\ \hline
               & $\ \ \ \ \ p\le0.94$\ \ \     & $\ \ \ >17.0\ $        \\
\ (0.40, 0.60)    & $\ \ \ \ \ 0.94<p<1.10$\ \ \  & $\ \ \ >100.0\times p-77.0\ $ \\
               & $\ \ \ \ \ p\ge1.10$\ \ \     & $\ \ \ >33.0\ $ \\ \hline
\ (0.60, 0.80)    & \ \ \ \ \ -  \ \ \            & $\ \ \ >17.0\ $ \\ \hline
\ (0.80, 0.93)    & \ \ \ \ \ -  \ \ \            & $\ \ \ >17.0\ $ \\ \hline
\end{tabular}
\label{tab:muonid}
\end{center}
\end{table}

\begin{figure*}[!htbp] 
\centering
\setlength{\belowcaptionskip}{-0.1cm}  
\includegraphics[height=6.3cm]{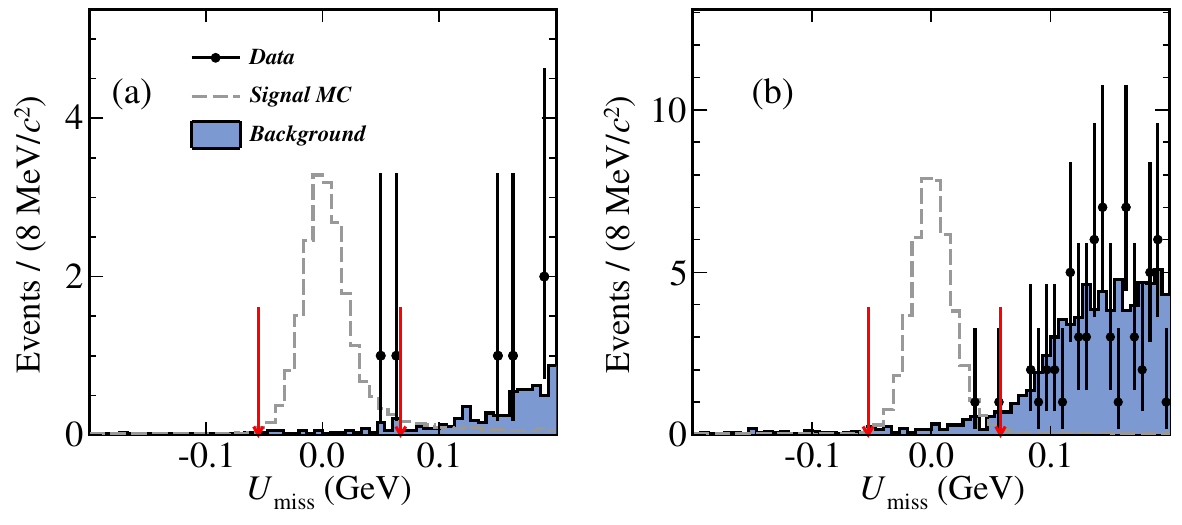}
\caption{\footnotesize
The $U_{\rm miss}$ distributions for the (a) $D^{*+}\to e^+\nu_e$ and (b) $D^{*+}\to \mu^+\nu_\mu$ candidates. The points with error bars are data combined from all energy points. The blue filled histogram is the simulated background derived from the inclusive MC sample. The dashed histogram is the signal MC events with arbitrarily normalization for visualization. The signal region lies inside the red arrows, and the sideband region lies outside.}  
\label{fig:dataumiss}
\end{figure*}

Neutrinos cannot be detected directly by the BESIII detector. However, their presence can be inferred in 
the process $e^+e^-\to D^{*+}D^{*-}$ by the kinematic
variable $U_{\rm miss} = E_{\rm miss} - |\vec{p}_{\rm miss}|$,
where $E_{\rm miss}$ is missing energy calculated by
$E_{\rm miss} = E_{\rm cm} - E_{D^{*-}} - E_{\ell^+}$, and
$\vec{p}_{\rm miss}$ is the momentum of the missing neutrino given by
$\vec{p}_{\rm miss} =  - \vec p_{D^{*-}}- \vec{p}_{\ell^+}$,
where $E_{\ell^+}$ and $\vec{p}_{\ell^+}$ are the energy and
momentum of $e^+$ or $\mu^+$ in the $e^+e^-$ CM frame, respectively.
A signal would manifest itself as an excess of events around $U_{\rm miss} = 0$. 
No significant signal is seen in the events passing the $D^{*+}\to \ell^+\nu_\ell$  selection, as can be seen  in Fig.~\ref{fig:dataumiss}. 

The $U_{\rm miss}$ signal and sideband regions are defined based on
the signal MC events. Fits to signal MC events
distributions are performed by using a double Gaussian function to model the signal shape and a third-order polynomial to model the background shape. The $U_{\rm miss}$ signal region is set to be three standard deviations around the fitted peaks, which correspond to $(-0.055, 0.067)$~GeV for $D^{*+}\to e^+\nu_e$ and
$(-0.053, 0.058)$~GeV for $D^{*+}\to \mu^+\nu_\mu$, respectively. The sideband region for both selections is defined to be $U_{\rm miss}$ within $[-0.2, 0.2]$~GeV and lying outside the signal region. 

The DT efficiencies for $D^{*+}\to\ell^+\nu_\ell$ events are calculated from the number of signal MC events falling into the signal regions divided by the number of generated signal MC events, in
which $D^{*+}$ decays to the $\ell^+\nu_\ell$ final state and $D^{*-}$ decays to the tag modes. The DT
efficiencies~($\epsilon^i_{\rm DT}$) for each tag mode at $\sqrt{s}=4.178$~GeV are listed in Table~\ref{tab:dteff}, and those for the other
energy points can be found in Appendix~\ref{app:DT}. From the DT efficiencies presented in this table, and the ST efficiencies listed in 
Table~\ref{tab:steff}, the effective averaged efficiencies for 
reconstructing the $D^{*+}\to \ell^+\nu_\ell$ decay,
$\bar \epsilon_{D^{*+}\to \ell^+\nu_\ell}$, are determined to be
$(80.90\pm1.46)\%$ for $D^{*+}\to e^+\nu_e $ and $(68.86\pm1.23)\%$ for
$D^{*+}\to \mu^+\nu_\mu $, after applying the muon PID systematic correction as described in
Sec.~\ref{sec:sys}. 

\begin{table}[htbp]
 \renewcommand\arraystretch{1.25}
  \centering
  \caption{The DT efficiencies of $D^{*+} \to \ell^+ \nu_\ell$ at 4.178~GeV.
    Uncertainties are statistical only.}
  \label{tab:dteff}
  \begin{tabular}{lcc}
    \hline
    \multirow{2}*{Tag mode} & \multicolumn{2}{c}{$\epsilon^i_{\rm DT}$ (\%)}\\
                            & $D^{*+} \to e^+ \nu_e$ & $D^{*+} \to \mu^+ \nu_\mu$\\ \hline
    $\bar{D}^{0} \to K^+\pi^-$          & $6.90 \pm0.06$ & $6.47 \pm0.06$\\
    $\bar{D}^{0}\to K^+\pi^-\pi^+\pi^-$ & $3.87 \pm0.07$ & $3.72 \pm0.07$\\
    $\bar{D}^{0}\to K^+\pi^-\pi^0$      & $4.10 \pm0.05$ & $3.91 \pm0.04$\\
    $D^- \to K^+\pi^-\pi^-$             & $17.38\pm0.09$ & $16.65\pm0.09$\\
    $D^- \to K^+K^-\pi^-$               & $13.31\pm0.11$ & $12.75\pm0.11$\\
    $D^- \to K^+\pi^-\pi^-\pi^0$        & $8.81 \pm0.09$ & $8.27 \pm0.08$\\
    $D^- \to K_S^0\pi^-$                & $16.97 \pm0.11$& $16.08 \pm0.11$\\
    $D^- \to K_S^0\pi^-\pi^0$           & $9.37 \pm0.12$ & $9.01 \pm0.12$\\
    $D^- \to K_S^0\pi^-\pi^+\pi^-$      & $9.88 \pm0.16$ & $9.54 \pm0.16$\\ \hline
  \end{tabular}
\end{table}

From counting, we determine $N^{\rm data}_{\rm SR}$, the number of events in the $U_{\rm miss}$ signal region in data, and  $N^{\rm data}_{\rm SB}$, the number in the data sideband region.  The corresonding numbers in the inclusive MC sample are $N^{\rm MC}_{\rm SR}$ and $N^{\rm MC}_{\rm SB}$. 
We estimate the number of background events in the signal region to be  $N^{\rm data}_{\rm SB}$ scaled by the factor ($N^{\rm MC}_{\rm SR}/N^{\rm MC}_{\rm SB}$).  The number of events in each region for data and MC simulation  are listed in Table~\ref{tab:brcalc}. 

\section{Upper limits of branching fractions}
The upper limits on the signal yields of the $D^{*+}\to e^+\nu_e $ and
$D^{*+}\to \mu^+\nu_\mu $ decays at the 90\% confidence level, $N_{\rm UL}$, are
calculated by using a frequentist method with an unbounded profile likelihood
treatment of systematic uncertainties, as implemented by the TROLKE package in
the ROOT software~\cite{Torlke}, where the number of the signal and background events are assumed to follow a
Poisson distribution, the detection efficiency is assumed to follow a Gaussian distribution, and the systematic uncertainty, which will be discussed below, is considered to be the standard deviation of the efficiency.  


Since the effective averaged efficiency $\bar \epsilon_{D^{*+}\to \ell^+\nu_\ell}$ has been considered in the determination of $N_{\rm UL}$ by the TROLKE package, 
the upper limits of the branching fractions are given by
\begin{equation}\label{eq:upbr}
  \mathcal{B}_{\rm UL} = \frac{N_{\rm UL}}{N^{\rm tot}_{\rm ST}}.
\end{equation}
Inserting the relevant quantities from Table~\ref{tab:brcalc} into Eq.~\ref{eq:upbr},
the upper limits of the branching fractions of $D^{*+}\to e^+\nu_e $ and
$D^{*+}\to \mu^+\nu_\mu$  are determined to be $1.1\times10^{-5}$ and
$4.1\times10^{-6}$, respectively, at the 90\% confidence level.

\begin{table*}[htbp]
 \renewcommand\arraystretch{1.25}
  \centering
\caption{Summary of the quantities used for branching fraction calculation.}
\label{tab:brcalc}
\begin{tabular}{lcccccccc}
\hline
Signal mode             & $N^{\rm tot}_{\rm ST}$         &$N^{\rm data}_{\rm SR}$  &$N^{\rm data}_{\rm SB}$ &$N^{\rm MC}_{\rm SR}$  &$N^{\rm MC}_{\rm SB}$ & $\bar\epsilon_{D^{*+}\to \ell^+\nu_\ell} (\%)$ & $N_{\rm UL}$ & $\mathcal{B}_{\rm UL}$ \\ \hline
$D^{*+}\to e^+\nu_e$    & \multirow{2}*{$516256\pm1870$} &$2$&$4$&$34$&$239$& $80.90\pm1.46$      & 5.9      & $1.1\times10^{-5}$  \\
$D^{*+}\to\mu^+\nu_\mu$ &                                &$1$&$61$&$182$&$2872$& $68.86\pm1.23$       & 2.2      & $4.3\times10^{-6}$  \\ \hline
\end{tabular}
\end{table*}

\section{SYSTEMATIC UNCERTAINTY}
\label{sec:sys}
Table~\ref{tab:Bf-syst-sum}
summarizes the assigned systematic uncertainties, which are discussed below. 

The systematic uncertainties associated with $e^\pm(\mu^\pm)$ tracking and PID efficiencies
are studied with a control sample of $e^+e^-\to\gamma e^+e^-(\mu^+\mu^-)$ events. The efficiencies determined from this control sample are used to correct the efficiencies measured in the 
signal MC samples in the two dimensions of momentum and $\cos\theta$ and the difference in result taken as the corresponding systematic uncertainty for the tracking and electron PID efficiency.   In the case of the muon PID, the systematic uncertainty is assigned by propagating the statistical
uncertainty associated with the control sample to the branching-fraction measurement .  

The uncertainty in the total number of ST $D_s^-$ mesons is assigned to
be 2.9\% by examining the changes in the fit yields when varying the signal
and  background shapes in
the fit. The nominal signal shape is replaced with an MC-simulated shape
convolved with a double-Gaussian function and the nominal background shape
with an ARGUS function~\cite{ARGUS}.

The uncertainties associated with the extra photon energy and extra charged track requirements are studied with samples of the light hadronic processes $e^+e^-\to K^+K^-\pi^+\pi^-$,
$\pi^+\pi^-\pi^+\pi^-$, $K^+K^-\pi^+\pi^-\pi^0$ and $\pi^+\pi^-\pi^+\pi^-\pi^0$.  The ratios of the average efficiencies of data to those of simulation are found to be  $1.008 \pm 0.001$ for the extra photon-energy requirement and $0.998 \pm 0.001$ for the extra charged-track requirement. We assign $0.8\%$ and $0.2\%$ as the systematic uncertainties from the extra photon energy and charged track requirements, respectively.

The uncertainty associated with the ST efficiency does not fully cancel. Because the ST efficiencies estimated with the inclusive and signal MC samples differ from each other due to different multiplicities, there can be a bias associated with the reconstruction of the  tag mode. We study the variation in 
tracking/PID efficiencies for different multiplicities, and take the combined
differences between data and MC simulation, 0.7\% for $D^{*+}\to e^+\nu_e$ and
0.8\% $D^{*+}\to\mu^+\nu_\mu$, as the corresponding tag-bias uncertainties. The efficiency for reconstructing the tag modes almost cancel with the DT
method, and the residual effects are referred to as tag bias.

The systematic uncertainty caused by the change of $f$ factor is estimated to be $4.1\%$ as a result of cross section uncertainty of $D^{*+}D^{*-}$, $D^{*-}D^{+}$,  $D^{0}D^{*-}\pi^{+}$, $D^{+}D^{*-}\pi^{0}$, even including $D^{*0}D^{*-}\pi^{+}$~\cite{D*D*pi} and $D^{*+}D^{*-}\pi^{0}$. Similarity, according to the isospin symmetry, the cross section of $D^{*+}D^{*-}\pi^{0}$ is set to be half of $D^{*0}D^{*-}\pi^{+}$. The uncertainty in the knowledge of the ST and DT efficiencies arising from the limited MC sample sizes, as shown in Table~\ref{tab:steff} and \ref{tab:dteff}, is evaluated to be 1.0\% for each signal decay mode.

\begin{table}[htbp]
 \renewcommand\arraystretch{1.25}
  \centering
  \caption{Relative systematic uncertainties (in \%) from each source.}
  \begin{tabular}{ccc}
\hline
Source &$D^{*+}\to e^+\nu_e$ & $D^{*+}\to\mu^+\nu_\mu$ \\ \hline
Tracking & 0.2 & 0.2 \\
PID      & 0.5 & 0.5 \\
$N^{\rm tot}_{\rm ST}$ & 2.9 & 2.9 \\
$E_{\rm{extra}~\gamma}^{\rm max}<0.3$~GeV  & 0.8 & 0.8 \\
$N_{\rm extra}^{\rm charge}=0$  & 0.2 & 0.2 \\
Tag bias           & 0.7 & 0.8 \\
$f$ factor       & 4.1 & 4.1 \\ 
MC sample sizes      & 1.0 & 1.0 \\ \hline
Total              & 5.3 & 5.3 \\ \hline
\end{tabular}
\label{tab:Bf-syst-sum}
\end{table}


\section{Conclusion} \label{CONLUSION}
Using a data sample corresponding to an integrated luminosity of
$6.32~\mathrm{fb}^{-1}$, taken at $\sqrt{s} = $ 4.178-4.226~GeV by the
BESIII detector, we search for the leptonic decays of $D^{*+} \to e^+\nu_e$ and
$D^{*+} \to \mu^+\nu_\mu$ for the first time. No significant signal is
observed. We set upper limits on the branching fractions of these decays, which are  $\mathcal{B}(D^{*+} \to e^+\nu_e)<1.1\times10^{-5}$ and
$\mathcal{B}(D^{*+} \to \mu^+\nu_\mu)<4.3\times 10^{-6}$ at the 90\% confidence
level. The larger data sets that are foreseen to be collected at
BESIII in the coming years~\cite{Ablikim:2019hff} will offer the opportunity to further improve the sensitivity to these decays.

\begin{acknowledgements}
\label{sec:acknowledgement}
\vspace{-0.4cm}

The BESIII collaboration thanks the staff of BEPCII and the IHEP computing center for their strong support. This work is supported in part by National Key R\&D Program of China under Contracts Nos. 2020YFA0406300, 2020YFA0406400; National Natural Science Foundation of China (NSFC) under Contracts Nos. 12035009, 11875170, 11875054, 11635010, 11735014, 11835012, 11935015, 11935016, 11935018, 11961141012, 12022510, 12025502, 12035013, 12192260, 12192261, 12192262, 12192263, 12192264, 12192265; the Chinese Academy of Sciences (CAS) Large-Scale Scientific Facility Program; Joint Large-Scale Scientific Facility Funds of the NSFC and CAS under Contract No. U1832207, U2032104; the CAS Center for Excellence in Particle Physics (CCEPP); 100 Talents Program of CAS; The Institute of Nuclear and Particle Physics (INPAC) and Shanghai Key Laboratory for Particle Physics and Cosmology; ERC under Contract No. 758462; European Union's Horizon 2020 research and innovation programme under Marie Sklodowska-Curie grant agreement under Contract No. 894790; German Research Foundation DFG under Contracts Nos. 443159800, 455635585, Collaborative Research Center CRC 1044, FOR5327, GRK 2149; Istituto Nazionale di Fisica Nucleare, Italy; Ministry of Development of Turkey under Contract No. DPT2006K-120470; National Science and Technology fund; National Science Research and Innovation Fund (NSRF) via the Program Management Unit for Human Resources \& Institutional Development, Research and Innovation under Contract No. B16F640076; Olle Engkvist Foundation under Contract No. 200-0605; STFC (United Kingdom); Suranaree University of Technology (SUT), Thailand Science Research and Innovation (TSRI), and National Science Research and Innovation Fund (NSRF) under Contract No. 160355; The Royal Society, UK under Contracts Nos. DH140054, DH160214; The Swedish Research Council; U. S. Department of Energy under Contract No. DE-FG02-05ER41374.
\end{acknowledgements}

\appendix
\section{ST yields and efficiencies}
\label{app:ST}
\begin{figure}[h]
  \centering
  \includegraphics[width=0.48\textwidth]{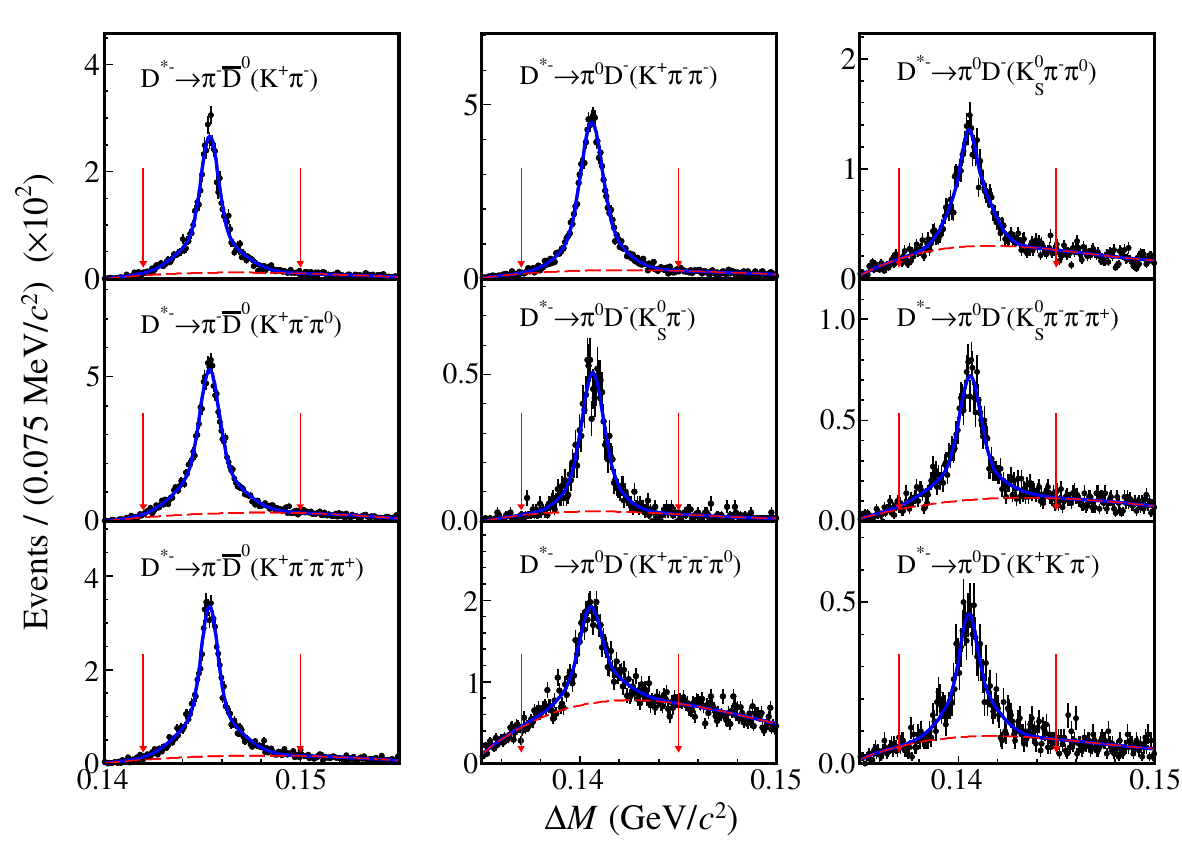}
  \caption{Fits to the $\Delta M$ distributions of the accepted ST $D^{*-}$
    candidates at $\sqrt{s}=4.189$~GeV. The points with error bars are data. The blue solid
    curves and red dashed curves represent the best fits and fitted combinatorial
    backgrounds, respectively. The pairs of red arrows indicate the $\Delta M$
    signal region.}
    \label{fig:styeild_4190}
\end{figure}

\begin{figure}[h]
  \centering
  \includegraphics[width=0.48\textwidth]{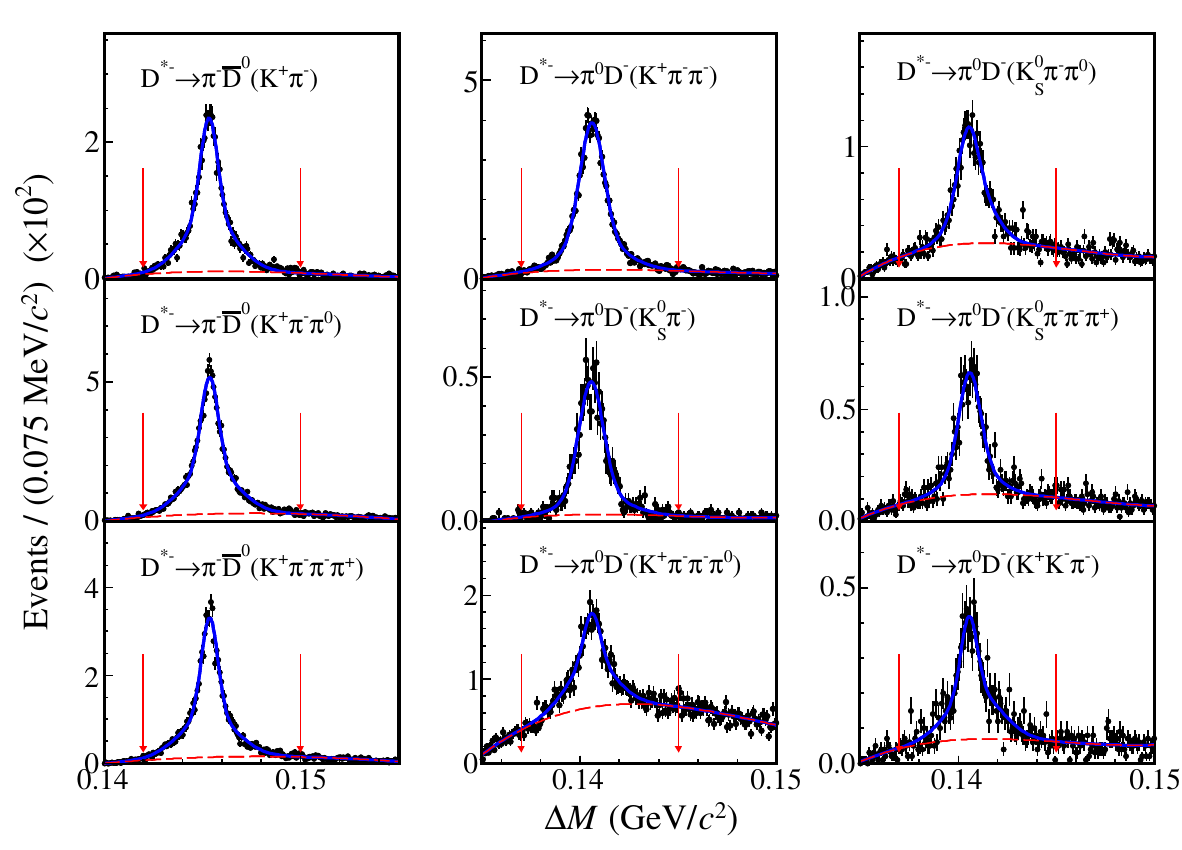}
  \caption{Fits to the $\Delta M$ distributions of the accepted ST $D^{*-}$
    candidates at $\sqrt{s}=4.199$~GeV. The points with error bars are data. The blue solid
    curves and red dashed curves represent the best fits and fitted combinatorial
    backgrounds, respectively. The pairs of red arrows indicate the $\Delta M$
    signal region.}
  \label{fig:styeild_4200}
\end{figure}

\begin{figure}[h]
  \centering
  \includegraphics[width=0.48\textwidth]{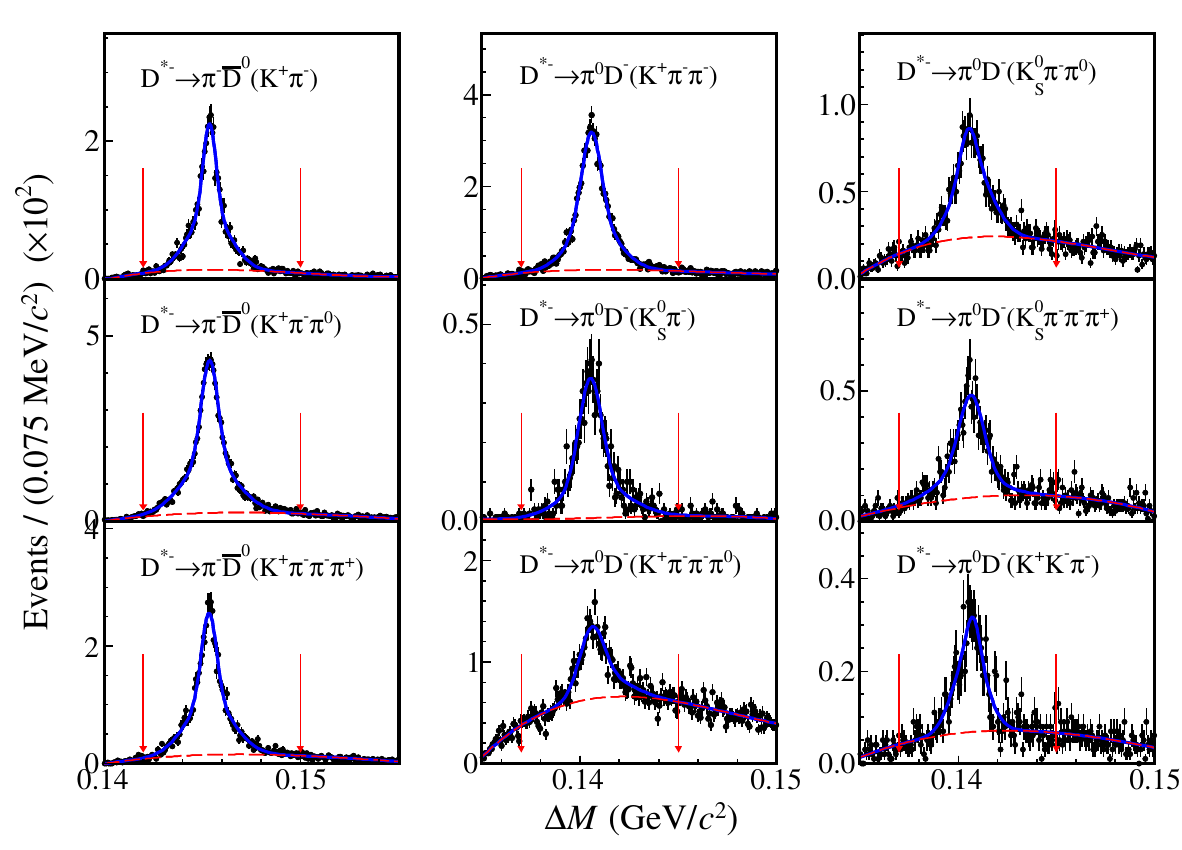}
  \caption{Fits to the $\Delta M$ distributions of the accepted ST $D^{*-}$
    candidates at $\sqrt{s}=4.209$~GeV. The points with error bars are data. The blue solid
    curves and red dashed curves represent the best fits and fitted combinatorial
    backgrounds, respectively. The pairs of red arrows indicate the $\Delta M$
    signal region.}
  \label{fig:styeild_4210}
\end{figure}

\begin{figure}[h]
  \centering
  \includegraphics[width=0.48\textwidth]{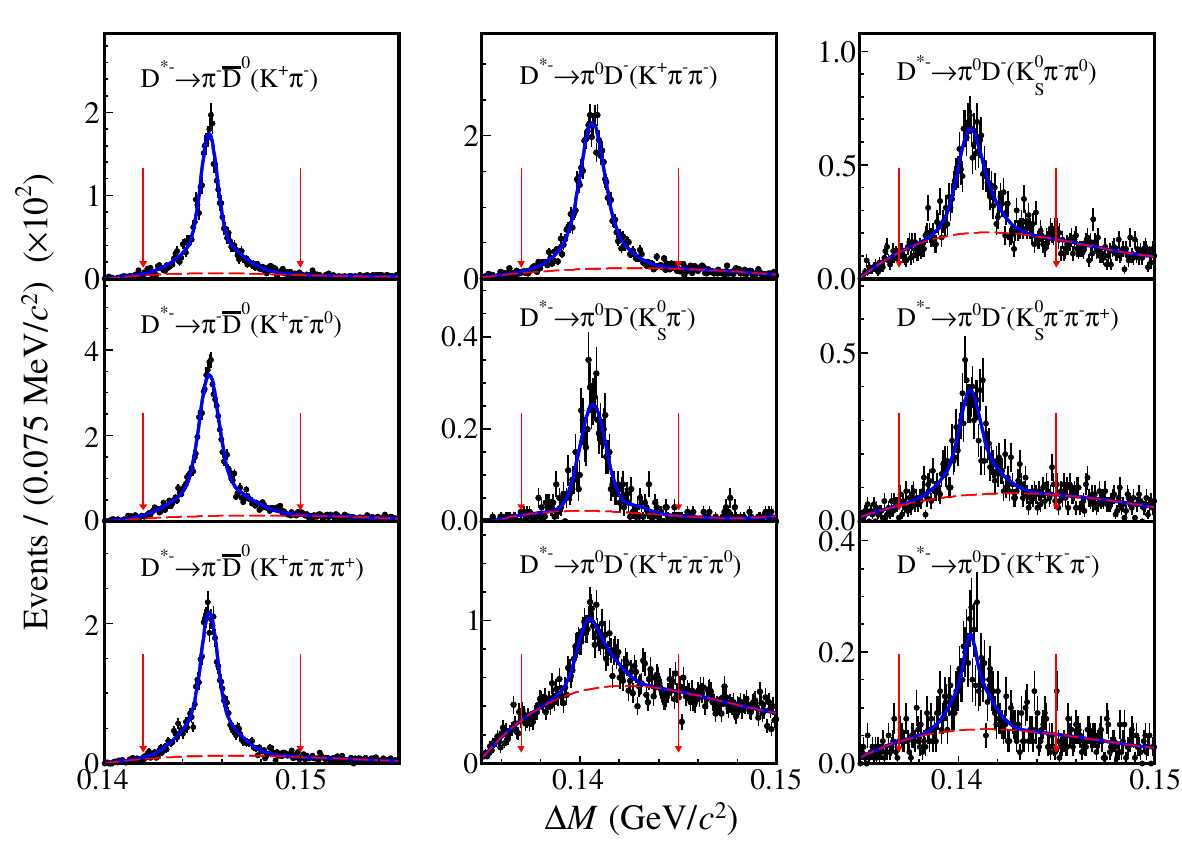}
  \caption{Fits to the $\Delta M$ distributions of the accepted ST $D^{*-}$
    candidates at $\sqrt{s}=4.219$~GeV. The points with error bars are data. The blue solid
    curves and red dashed curves represent the best fits and fitted combinatorial
    backgrounds, respectively. The pairs of red arrows indicate the $\Delta M$
    signal region.}
  \label{fig:styeild_4220}
\end{figure}

\begin{figure}[h]
  \centering
  \includegraphics[width=0.48\textwidth]{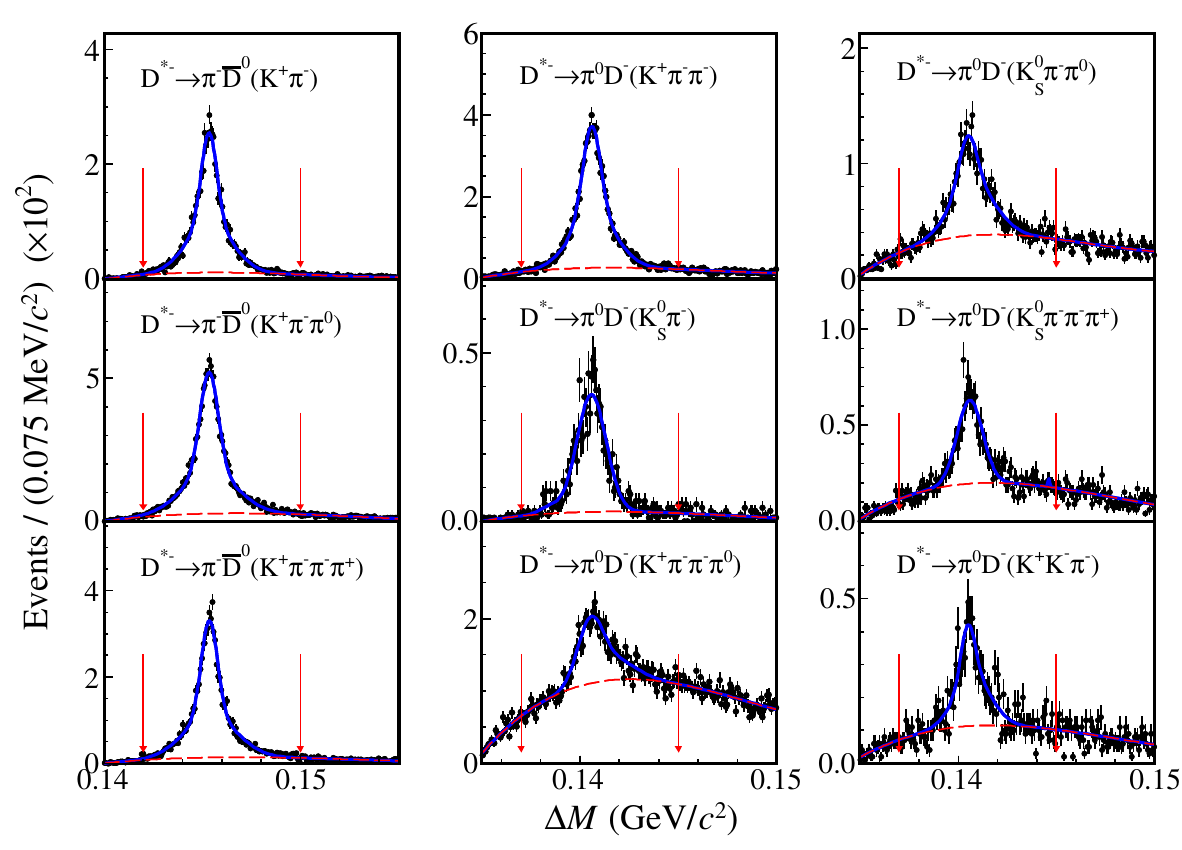}
  \caption{Fits to the $\Delta M$ distributions of the accepted ST $D^{*-}$
    candidates at $\sqrt{s}=4.226$~GeV. The points with error bars are data. The blue solid
    curves and red dashed curves represent the best fits and fitted combinatorial
    backgrounds, respectively. The pairs of red arrows indicate the $\Delta M$
    signal region.}
  \label{fig:styeild_4230}
\end{figure}

Tables~\ref{tab:steff_scan}~and~\ref{tab:styields_scan} summarize the ST efficiencies and the ST yields in data at $\sqrt{s}=4.189-4.226$~GeV, respectively. 
Figures~\ref{fig:styeild_4190}-\ref{fig:styeild_4230} show the fits to the
$\Delta M$ distributions of the accepted ST $D^{*-}$ candidates at these energy points.

\begin{table*}[h]
  \renewcommand\arraystretch{1.25}
  \centering
  \caption{ST efficiencies at $\sqrt{s}=4.189-4.226$~GeV. The uncertainties are statistical
    only.}
  \label{tab:steff_scan}
  \begin{tabular}{lcccccc}
    \hline
    \multirow{2}*{Tag mode} & \multicolumn{5}{c}{$\epsilon^i_{\rm ST}$ (\%)}\\
                         & 4.189          & 4.199          & 4.209          & 4.219         & 4.226 \\ \hline
$\bar{D}^{0}\to K^{-}\pi^{+}$ & $8.66\pm0.23$ & $7.89\pm0.17$ & $8.15\pm0.22$ & $5.50\pm0.17$ & $5.33\pm0.20$ \\
$\bar{D}^{0}\to K^{-}\pi^{+}\pi^{+}\pi^{-}$ & $3.98\pm0.11$ & $4.73\pm0.10$ & $4.36\pm0.10$ & $3.30\pm0.12$ & $2.97\pm0.08$ \\
$\bar{D}^{0}\to K^{-}\pi^{+}\pi^{0}$ & $4.25\pm0.10$ & $4.86\pm0.09$ & $4.56\pm0.09$ & $3.62\pm0.10$ & $3.53\pm0.33$ \\
$D^{+}\to K^{-}\pi^{+}\pi^{+}$ & $12.28\pm0.17$ & $14.06\pm0.23$ & $13.13\pm0.22$ & $9.26\pm0.23$ & $8.38\pm0.66$ \\
$D^{+}\to K^{+}K^{-}\pi^{+}$ & $8.88\pm2.43$ & $10.28\pm0.61$ & $11.22\pm1.48$ & $6.71\pm0.59$ & $5.72\pm0.55$ \\
$D^{+}\to K^{-}\pi^{+}\pi^{+}\pi^{0}$ & $6.41\pm0.44$ & $6.73\pm0.32$ & $5.30\pm0.31$ & $5.65\pm0.50$ & $4.04\pm0.29$ \\
$D^{+}\to K^{0}_{s}\pi^{+}$ & $12.09\pm0.56$ & $13.09\pm0.50$ & $12.92\pm0.72$ & $9.45\pm0.74$ & $8.75\pm0.98$ \\
$D^{+}\to K^{0}_{s}\pi^{+}\pi^{0}$ & $7.44\pm0.48$ & $7.22\pm0.30$ & $6.59\pm0.32$ & $3.71\pm0.23$ & $4.12\pm0.25$ \\
$D^{+}\to K^{0}_{s}\pi^{+}\pi^{+}\pi^{-}$ & $7.67\pm0.46$ & $8.08\pm0.37$ & $8.27\pm0.73$ & $5.54\pm0.66$ & $3.91\pm0.31$

\\\hline
\end{tabular}
\end{table*}

\begin{table*}[h]
  \renewcommand\arraystretch{1.25}
  \centering
  \caption{ST yields in data at $\sqrt{s}=4.189-4.226$~GeV. Uncertainties are statistical only.}
  \label{tab:styields_scan}
  \begin{tabular}{lcccccc}
    \hline
    \multirow{2}*{Tag mode} & \multicolumn{5}{c}{$N^{i}_{\rm ST}$}\\
                                          & 4.189         & 4.199         & 4.209         & 4.219        & 4.226 \\ \hline
    $\bar{D}^{0}\to K^{-}\pi^{+}$               & $6070\pm197$  & $5546\pm147$  & $4264\pm109$  & $3644\pm149$ & $5402\pm144$ \\
    $\bar{D}^{0}\to K^{-}\pi^{+}\pi^{+}\pi^{-}$ & $7626\pm214$  & $7444\pm207$  & $5511\pm144$  & $4632\pm125$ & $7243\pm187$ \\
    $\bar{D}^{0}\to K^{-}\pi^{+}\pi^{0}$        & $13081\pm342$ & $12143\pm443$ & $10059\pm232$ & $8316\pm309$ & $12004\pm262$ \\
    $D^{+}\to K^{-}\pi^{+}\pi^{+}$        & $10675\pm210$ & $9587\pm308$  & $7293\pm210$  & $5495\pm177$ & $8195\pm142$ \\
    $D^{+}\to K^{+}K^{-}\pi^{+}$          & $938\pm76$    & $899\pm85$    & $560\pm42$    & $370\pm126$  & $645\pm56$ \\
    $D^{+}\to K^{-}\pi^{+}\pi^{+}\pi^{0}$ & $3180\pm313$  & $1864\pm212$  & $2271\pm147$  & $1302\pm206$ & $2612\pm256$\\
    $D^{+}\to K^0_s\pi^{+}$               & $1120\pm50$   & $1151\pm84$   & $978\pm80$    & $546\pm32$   & $903\pm128$ \\
    $D^{+}\to K^0_s\pi^{+}\pi^{0}$        & $2590\pm121$  & $2162\pm82$   & $1625\pm79$   & $1175\pm69$  & $2300\pm152$ \\
    $D^{+}\to K^0_s\pi^{+}\pi^{+}\pi^{-}$ & $1664\pm281$  & $1234\pm88$   & $1097\pm160$  & $798\pm78$   & $951\pm57$ \\\hline
    Sum                                   & $46944\pm667$ & $43123\pm655$ & $33251\pm438$ & $26278\pm445$& $40255\pm472$\\\hline
  \end{tabular}
\end{table*}


\section{DT efficiencies}
\label{app:DT}
Tables~\ref{tab:dteff_e_scan}~and~\ref{tab:dteff_mu_scan} summarize the DT
efficiencies of $D^{*+}\to e^+ \nu_e$ and $D^{*+}\to \mu^+ \nu_\mu$ at
$\sqrt{s}=4.189-4.226$~GeV, respectively.
\begin{table*}[htbp]
  \renewcommand\arraystretch{1.25}
  \centering
  \caption{DT efficiencies of $D^{*+}\to e^+ \nu_e$ at $\sqrt{s}=4.189-4.226$~GeV.}
  \label{tab:dteff_e_scan}
  \begin{tabular}{lcccccc}
    \hline
    \multirow{2}*{Tag mode} & \multicolumn{5}{c}{$\epsilon^i_{{\rm ST},D^{*+}\to e^+\nu_e}$}\\
                                          & 4.189          & 4.199          & 4.209          & 4.219          & 4.226 \\ \hline
    $\bar{D}^{0}\to K^{-}\pi^{+}$               & $7.20\pm0.06$  & $7.52\pm0.06$  & $6.99\pm0.06$  & $6.67\pm0.06$  & $6.50\pm0.06$ \\
    $\bar{D}^{0}\to K^{-}\pi^{+}\pi^{+}\pi^{-}$ & $4.12\pm0.07$  & $4.29\pm0.07$  & $3.99\pm0.07$  & $3.91\pm0.07$  & $3.83\pm0.07$ \\
    $\bar{D}^{0}\to K^{-}\pi^{+}\pi^{0}$        & $4.22\pm0.05$  & $4.39\pm0.05$  & $4.25\pm0.05$  & $4.05\pm0.05$  & $3.86\pm0.04$ \\
    $D^{+}\to K^{-}\pi^{+}\pi^{+}$        & $17.34\pm0.09$ & $16.93\pm0.09$ & $15.68\pm0.09$ & $14.38\pm0.08$ & $12.66\pm0.08$ \\
    $D^{+}\to K^{+}K^{-}\pi^{+}$          & $13.31\pm0.11$ & $13.11\pm0.11$ & $11.94\pm0.10$ & $11.01\pm0.10$ & $9.74\pm0.09$ \\
    $D^{+}\to K^{-}\pi^{+}\pi^{+}\pi^{0}$ & $8.70\pm0.08$  & $8.65\pm0.08$  & $7.90\pm0.08$  & $7.28\pm0.08$  & $6.87\pm0.08$ \\
    $D^{+}\to K^0_s\pi^{+}$               & $16.85\pm0.09$ & $16.59\pm0.09$ & $15.13\pm0.09$ & $13.76\pm0.08$ & $12.22\pm0.08$ \\
    $D^{+}\to K^0_s\pi^{+}\pi^{0}$        & $9.20\pm0.07$  & $9.18\pm0.07$  & $8.46\pm0.07$  & $7.79\pm0.06$  & $7.12\pm0.06$ \\
    $D^{+}\to K^0_s\pi^{+}\pi^{+}\pi^{-}$ & $8.38\pm0.09$  & $8.32\pm0.09$  & $7.77\pm0.09$  & $7.03\pm0.08$  & $6.69\pm0.08$ 
    \\\hline
  \end{tabular}
\end{table*}

\begin{table*}[htbp]
  \renewcommand\arraystretch{1.25}
  \centering
  \caption{DT efficiencies of $D^{*+}\to \mu^+ \nu_\mu$ at $\sqrt{s}=4.189-4.226$~GeV.}
  \label{tab:dteff_mu_scan}
  \begin{tabular}{lcccccc}
    \hline
    \multirow{2}*{Tag mode} & \multicolumn{5}{c}{$\epsilon^i_{{\rm ST},D^{*+}\to \mu^+\nu_\mu}$}\\
                                          & 4.189          & 4.199          & 4.209          & 4.219          & 4.226 \\ \hline
    $\bar{D}^{0}\to K^{-}\pi^{+}$               & $6.74\pm0.06$  & $7.00\pm0.06$  & $6.64\pm0.06$  & $6.27\pm0.06$  & $6.02\pm0.05$ \\
    $\bar{D}^{0}\to K^{-}\pi^{+}\pi^{+}\pi^{-}$ & $3.90\pm0.07$  & $4.12\pm0.07$  & $4.00\pm0.07$  & $3.65\pm0.07$  & $3.59\pm0.07$ \\
    $\bar{D}^{0}\to K^{-}\pi^{+}\pi^{0}$        & $3.87\pm0.04$  & $4.15\pm0.05$  & $3.93\pm0.04$  & $3.67\pm0.04$  & $3.57\pm0.04$ \\
    $D^{+}\to K^{-}\pi^{+}\pi^{+}$        & $16.47\pm0.09$ & $16.03\pm0.09$ & $15.03\pm0.09$ & $13.49\pm0.08$ & $11.98\pm0.08$ \\
    $D^{+}\to K^{+}K^{-}\pi^{+}$          & $12.49\pm0.11$ & $12.38\pm0.11$ & $11.54\pm0.10$ & $10.34\pm0.10$ & $9.15\pm0.09$ \\
    $D^{+}\to K^{-}\pi^{+}\pi^{+}\pi^{0}$ & $8.23\pm0.08$  & $7.99\pm0.08$  & $7.42\pm0.08$  & $7.21\pm0.08$  & $6.36\pm0.07$ \\
    $D^{+}\to K^0_s\pi^{+}$               & $16.12\pm0.09$ & $15.69\pm0.09$ & $14.54\pm0.09$ & $13.18\pm0.08$ & $11.69\pm0.08$ \\
    $D^{+}\to K^0_s\pi^{+}\pi^{0}$        & $8.75\pm0.07$  & $8.69\pm0.07$  & $7.94\pm0.06$  & $7.30\pm0.06$  & $6.69\pm0.06$ \\
    $D^{+}\to K^0_s\pi^{+}\pi^{+}\pi^{-}$ & $7.95\pm0.09$  & $7.96\pm0.09$  & $7.12\pm0.08$  & $6.73\pm0.08$  & $6.16\pm0.08$ 
    \\\hline
  \end{tabular}
\end{table*}

\end{document}